# Direct Measurement of the Electron Beam Intensity Profile via Carbon Nanotube Tomography

Matthew Zotta[1], Sharadh Jois[1], Prathamesh Dhakras[1], Miguel Rodriguez[1], Ji Ung Lee[1]*

[1]College of Nanoscale Science and Engineering, SUNY Polytechnic Institute, Albany, NY.

Correspondence to: *jlee1@sunypoly.edu

**Abstract:** Electron microscopes have been improved to achieve ever smaller beam spots, a key parameter that determines the instrument's resolution. The techniques to measure the size of the beam, however, have not progressed to the same degree. There is an on-going need to develop detection technologies that can potentially characterize the smallest electron beam to further improve the capabilities of these instruments. We report on a new electron beam detector based on a single-walled carbon nanotube. Nanotubes are atomically smooth, have a well-defined diameter that is similar in size to the finest electron probes and can be used to directly measure the beam profile. By scanning the beam at different angles to the nanotube, we can accurately determine the spatial profile of an electron beam by applying tomographic reconstruction techniques.

**Main Text:** Scanning electron microscopes (SEMs) and scanning transmission electron microscopes (STEMs) are critical to advancing nearly all scientific disciplines including biological sciences, physical sciences, and engineering (*1-3*). These tools scan a focused beam of electrons across a sample plane to image objects down to nanometer and sub-nanometer scales. Determining the size and shape of the electron beam at the sample surface can help quantify and improve the resolution limits of these tools (*4,5*).

The methods used to measure the electron beam profile, however, have not kept up with the capabilities of the instruments. A wide range of techniques is available to measure and determine the electron intensity profile (*5-17*), with verifiable resolution down to only several nanometers while the actual beam is believed to be less than a nanometer in high-resolution instruments.

The most common technique to measure the beam profile uses a knife-edge (*10-17*). In the knife-edge (KE) technique, an electron beam scans over an edge, and one measures either the transmitted (*10-16*) or emitted (*17*) signal as a function of position. The KE technique, however, becomes increasingly unreliable as the beam spot approaches the nanometer scale because there is no ideal specimen with an atomically sharp edge (*6-8, 11,13,14*). Furthermore, transmission or scattering off the KE corner obscures the beam profile (*6-8, 11,12,16*), requiring assumptions to be made of the beam profile for its measurement such as a Gaussian distribution.

Here, we use a single-walled carbon nanotube (SWCNT) as the smallest detector that can directly profile the beam. We accomplish this by measuring the current in the nanotube as the electron beam scans across it. We place the SWCNT device shown in Figure 1 inside an SEM equipped with nanoprobes to initially measure the distribution of electrons in one scan direction. By scanning the beam at different angles to the nanotube, we reconstruct the beam profile in a manner similar to medical reconstruction in X-ray computed tomography (CT) (*18,19*).

A SWCNT has the advantage of being atomically smooth and its diameter can be 0.4 nm or smaller. (*20*-22) While these ultra-small SWCNTs can potentially resolve even the smallest electron beam, the results we report here use larger diameter nanotubes of 1.5 ~ 1.9 nm (*23*) and sub-optimal imaging conditions to emphasize the reconstruction technique that we have developed. SWCNTs have negligible interaction volume which greatly reduces the scattering of an incident electron beam. To decrease the background signal further, we place the nanotubes over a trench (Fig. 1A). Most electron detectors are based on p-n diodes, which we have created with semiconducting nanotubes using buried split gates (Fig. 1B) (*24*). However, as we discuss below, the signal generated in the nanotube is plasmonic in origin. Therefore, we choose to report on metallic nanotubes which enhance plasmonic excitations (Fig. S1). In this case, the buried gates were only used to ground and create a symmetric device.

To understand the mechanism of the current generation in the nanotubes, we measure currents from both electrodes that contact the nanotube. If the beam excites electrons and holes in the nanotube, the source and drain contacts will measure equal and opposite currents. Instead, in both metallic and semiconducting nanotubes, we observe a sharp current consistent with electrons being injected into the nanotube from both contacts. These currents are in response to plasmons that decay in both directions from the point of impact and emit secondary electrons (*25*).

In measuring the currents from the two contacts, a small voltage difference of few microvolts results from the preamplifiers. This difference creates a DC background current of equal and opposite values on the source and drain contacts. By summing the two currents, we can eliminate this DC current. In addition, the nanotube collects the secondary electrons from the substrate at the bottom of the trench. It is known that secondary electrons resulting from an electron beam incident normal to a surface are emitted in a cosine distribution (*26*). In supplementary materials, we describe in detail how we subtract these background currents to extract the signal from the beam. (Fig. S2)

We show a typical measured profile in Fig. 2C for a focused beam at 2KeV. The resulting peak with full width at half max (FWHM) of 14 nm is a measure of the spot size and is consistent with the conditions we used for the SEM. Beam size is affected by a number of instrument and environmental parameters. The former includes beam energy, working distance, focus and astigmatism settings; the latter includes stray electromagnetic fields and acoustic vibrations. The rest of the discussion is based on a working distance of around 7 mm and a beam energy of 2.0 keV. The relatively long distance, which is not optimal for imaging, allows the placement of the nanoprobes onto the bondpads.

To demonstrate the reconstruction technique, we introduce a known astigmatism in the beam that results in a skewed, dumbbell-shaped beam, which we verified qualitatively by imaging well-defined gold particles of approximately 30 nm diameter. (Fig S3) The astigmatic beam is scanned perpendicular to the nanotube for a set path angle. The current is measured as a function of time and translated into spatial units by multiplying it with the beam speed. The correlation of temporal data and spatial data is used to convert the current measurements as a function of beam position relative to the nanotube. To form a tomographic reconstruction of the beam, we rotate the nanotube about the fixed beam. (Fig. S4) For each angle, the beam path must also be rotated such that the beam is scanned perpendicularly to the new nanotube orientation. At each measurement angle, the beam shape is adjusted to have the same astigmatism between each measurement as illustrated in Fig. S5.

Figure 3 shows results from scan angles ranging from 0º to 180º in steps of 18º for a total number of 11 measurements. The astigmatism is apparent as the FWHM varies sinusoidally as a function of scan angle. This is consistent with what would be expected for our dumbbell-shaped beam. We note that the FWHM for 0 and 180 degrees should be equivalent as these represent scanning the beam along the same path in the forward and reverse direction, respectively. The nanotube is known to contaminate and thereby swell which accounts for the difference in FWHM observed in the data. We present the effects of contamination and steps we took to minize its effects in supplementary materials (Fig S6).

The important realization is that each spatial profile is equivalent to a single projection of a CT scan from a parallel x-ray source (*18,19*). Here, the projection is a measure of the integrated beam profile perpendicular to the scan direction for one angle which generates the Radon transform (*27*). The collection of projections forms an image which is typically plotted as a sinogram shown in Fig. 4A. As in the CT reconstruction, the back-projection is used to begin the process of reconstructing the beam profile. To reconstruct the beam profile, we work in the Fourier domain supplementary materials. According to the Fourier slice theorem, the 1D Fourier Transform (FT) of the projection is a slice through the 2D FT of the beam profile at the same angle. The collection of each FT taken at different angles form a 2D function in the frequency domain. A direct inverse FT of the resulting function cannot be taken to reconstruct the beam in the spatial domain because the center is now oversampled, which results in significant blurring in the reconstruction. The reconstructed beam profile ($I'(x,y)$), however, can be formed by using a filtered back projection to attenuate low frequecy contributions. We use the 'Ram-Lak' filter shown in Fig. S7. Figure 4B shows the reconstruction of the electron beam profile using the filtered back projection technique.

In summary, we demonstrate a tomographic reconstruction technique to characterize the electron beam intensity profile with nanometer resolution using a SWCNT device as a direct detector. If smaller diameter nanotubes are implimented, the resolving limit of this technique can be pushed to characterize even the smallest electron and ion beams with diameter ≤ 1nm. Carbon build up on the nanotube can be reduced by mounting the device on a heated stage at 500°C. Active vibration cancellation system can dampen low frequency mechanical vibrations up to 1Hz. External electromagnetic noise cancellation will improve the stability of the beam. Exploring non-local and THz techniques could increase the resolution of the captured plasmonic signal. This new method opens the door for a more comprehensive measurement of electron beam interaction with nanostructured materials in general and for the electron beam profiling in electron microscopes in particular

**Acknowledgments:** The authors would like to thank Eric Lifshin and Bradley Thiel for useful discussions.

**Funding:** This research was supported by DTRA/NRL Grant No. N00173-14-1-G017, NSF Grant No. ECCS 1606659, and NSF AIR Grant No. 1127200.

**Supplementary Materials:**

Materials and Methods

Supplementary Text

Figures S1-S8

Tables S1-S2

References (*28-30*)

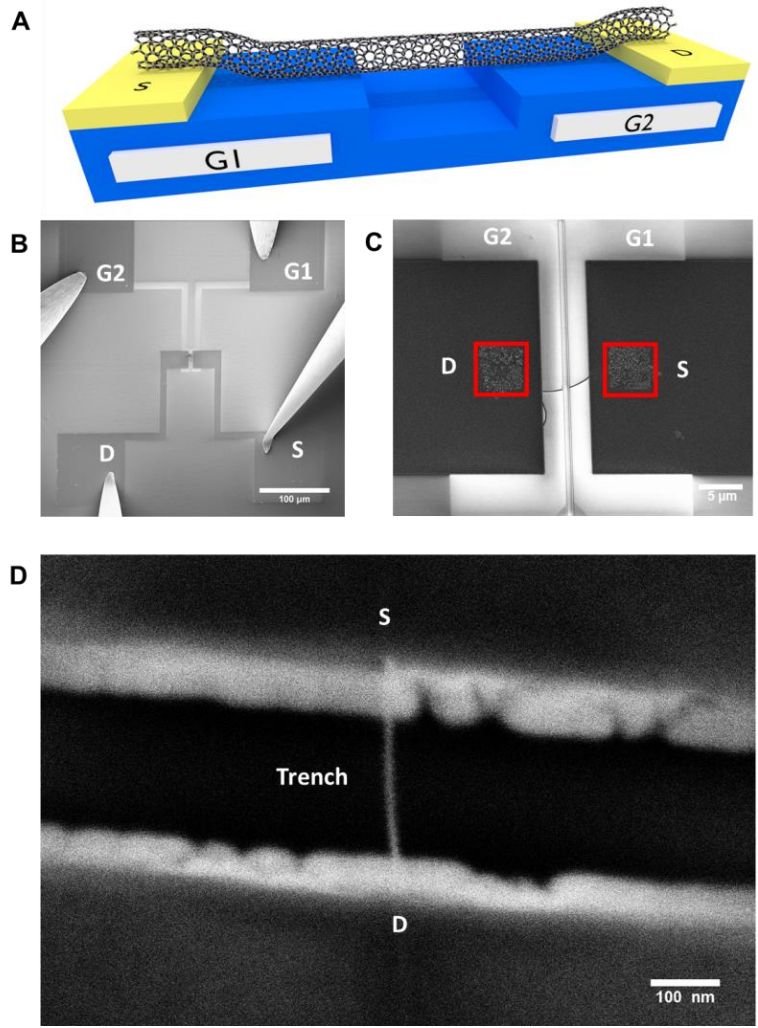

**Fig. 1.** (**A**) Device schematic showing the SWCNT electrically contacted to the source and drain and traversing a trench region in the dielectric material (SiO2). The SWCNT is fully suspended in the trench region. (**B**) An SEM image shows the source (S) and drain (D) nanoprobes landing on contact pads to measure current at both ends of the SWCNT as well as gate 1 (G1) and gate 2 (G2) nanoprobes for biasing the buried gates to dope semiconducting SWCNTs. The catalyst regions to grow the nanotubes are visible on the source and drain contacts (red outline) (**C**) An SEM image shows the location of the nanotube on the device. G1 and G2 are biased at -10V to increase the image contrast for locating the SWCNT. (**D**) A high magnification SEM image showing the SWCNT suspended in the trench region. Details of the device fabrication are presented in supplementary materials.

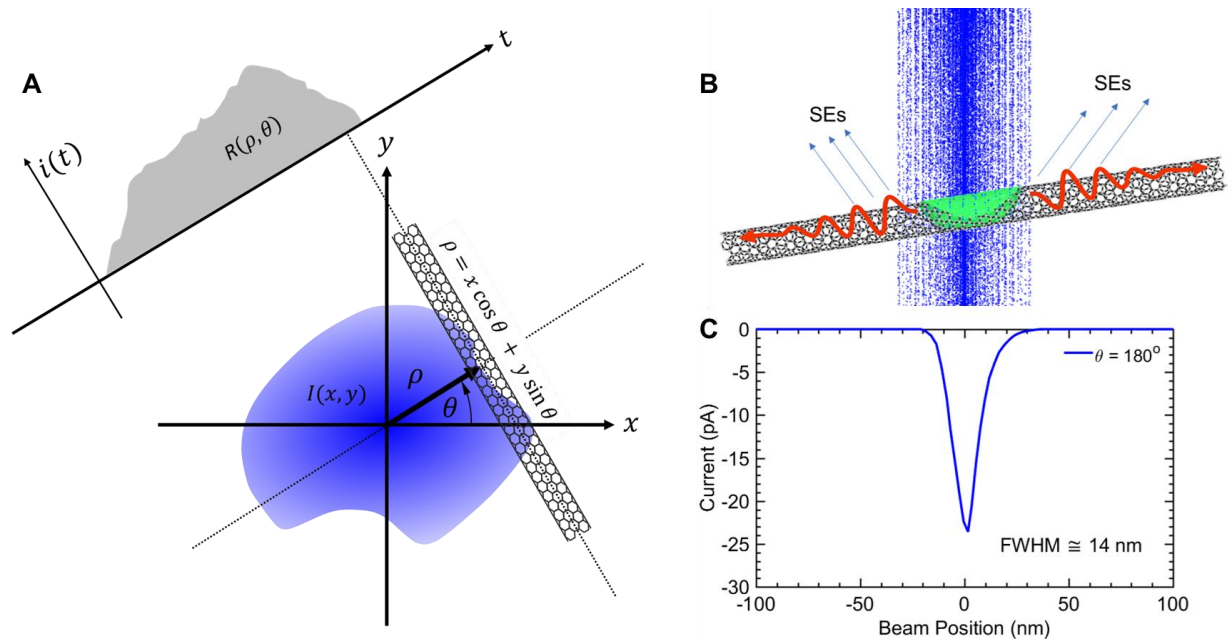

**Fig. 2.** (**A**) Schematic showing the electron beam intensity $I(x,y)$ as it crosses the nanotube in a perpendicular path along $\rho$. The current $i(t)$ in the nanotube is monitored over time as the beam scans, which generates the Radon transform for an angle $\theta$. A detailed explanation of the mathematics behind this schematic is given in supplementary materials. (**B**) When the beam strikes the nanotube, plasmons are generated and decay, ejecting secondary electrons into the vacuum. The plasmon signal can be directly correlated to the intensity of the electron beam. Electrons are injected into the nanotube from both contacts in response to the secondary electron emission. (**C**) The plasmon peak measured as a function of time can be converted to spatial data using the cycle time (seconds/line scan) and the path length of the line scan (nm).

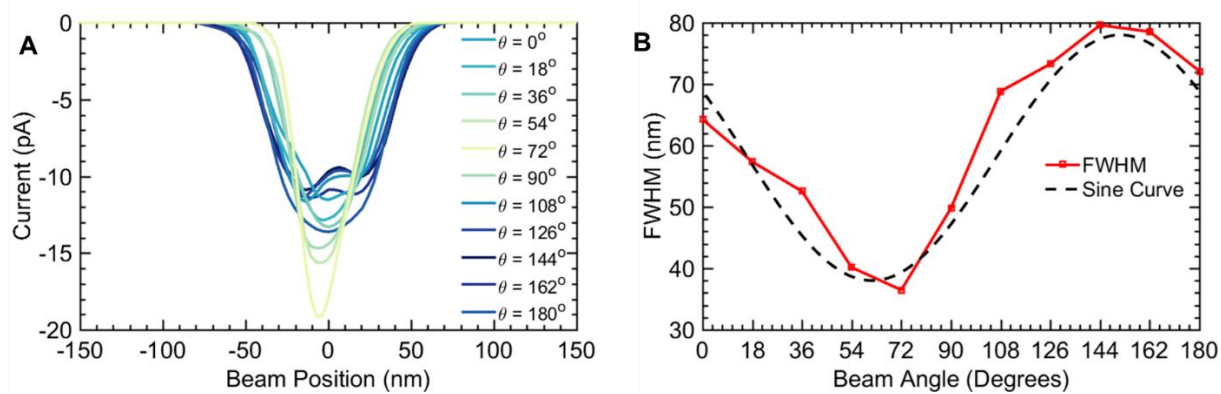

**Fig. 3.** (**A**) Transient measurement data of an astigmatic beam with the peak isolated from the background as discussed in supplementary materials. The device is rotated through 180 degrees in 18 degree increments for a total of 11 measurements. (**B**) A plot of the FWHM of each measurement. A sine curve can be seen to fit the data as the carbon nanotube is rotated through 180 degrees. This is expected for our dumbbell-shaped beam which was reproduced by setting the focus and astigmatism for each angle as laid out in Fig. S5.

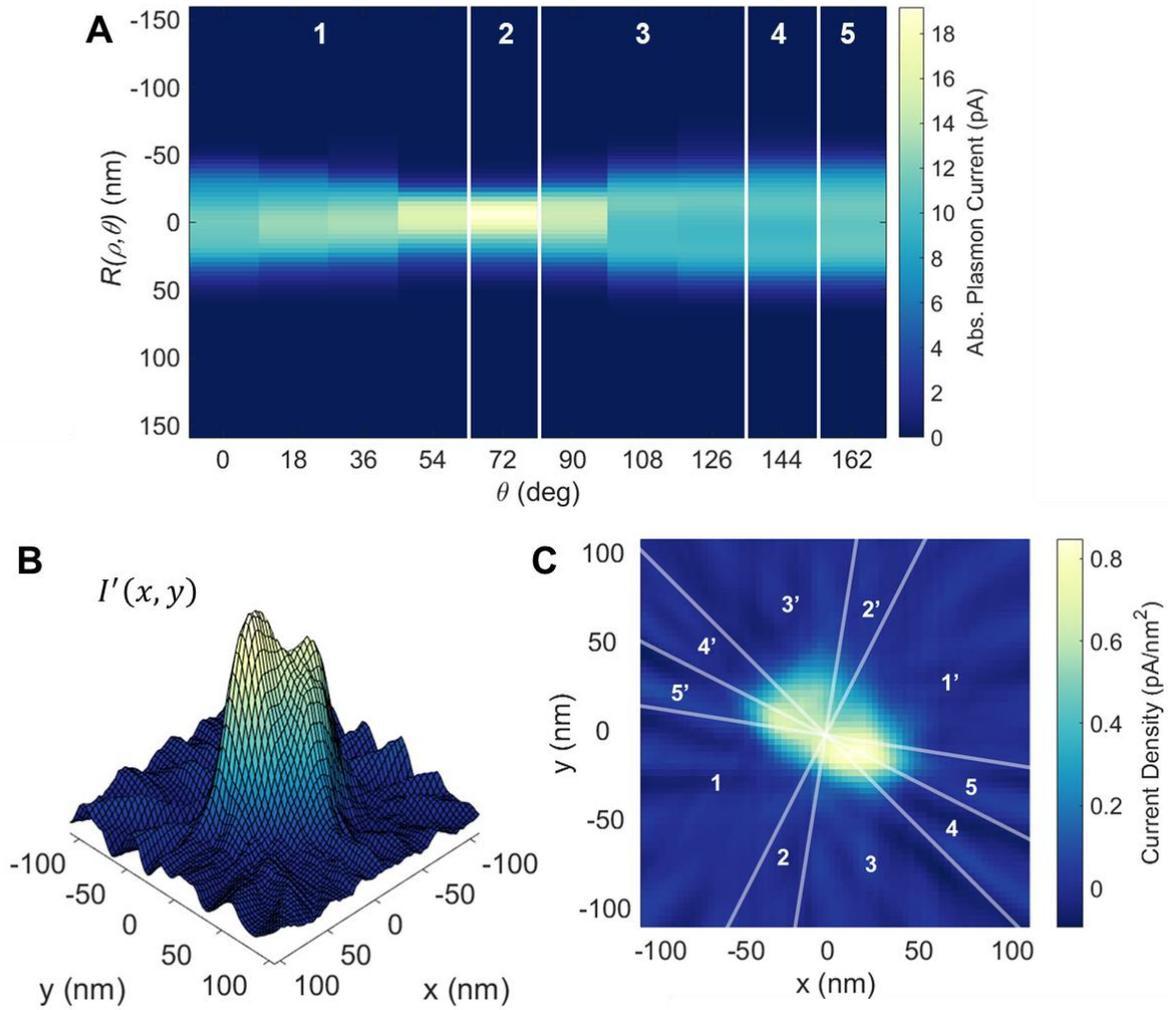

**Fig. 4**. (**A**) The transient data shown in Fig. 3A is displayed as a sinogram. The sinogram is a concise way to view the full Radon transform of the electron beam. An astigmatism has been intentionally applied to the electron beam resulting in two humps seen most clearly in the 144º beam profile. (**B**) A rotated and (**C**) top down view of the reconstructed electron beam distribution $I'(x, y)$ from the sinogram data. These are current density maps of the beam based on the measured beam current in a Faraday cup which was 220 pA. This was accomplished by normalizing the reconstruction to sum to unity and then multiplying each pixel by the beam current. The regions of the beam in (**C**) are numbered to correspond with the sinogram data in (**A**). A 'prime' indicates the compliment of the corresponding numbered region whose Radon transform is the mirrored profile of those in (**A**).

# Supplementary Materials for

## Direct Measurement of the Electron Beam Intensity Profile via Carbon Nanotube Tomography

Matthew Zotta, Sharadh Jois, Prathamesh Dhakras, Miguel Rodriguez, Ji Ung Lee.

Correspondence to: jlee1@sunypoly.edu

**This PDF file includes:**
Materials and Methods
Supplementary Text
Figs. S1 to S8
Tables S1 to S2

**Contents**

# 1. Materials and Methods

## 1.1. Device Fabrication

The details of the device fabrication have been discussed extensively elsewhere (*23*). Briefly, we use SUNY Poly's 300mm Si fabrication facility to fabricate the wafers. We begin by growing 100nm of thermal oxide on Si wafers. Next, we deposit 100nm of polysilicon using a chemical vapor deposition (CVD) process. The polysilicon is made highly conductive through phosphorous implantation and activation anneal. The implant dopant density is $10^{19}$-$10^{20}$ cm$^{-3}$. After implantation, the polysilicon gates are defined using photolithography and reactive ion etching (RIE) techniques. Hundreds of devices are fabricated at the same time with separation between the gates that vary between 0.1-20 µm. We next deposit 300nm of SiO2 using a plasma-enhanced chemical vapor deposition (PECVD) process with tetra-ethyl-ortho-silicate (TEOS) and ozone precursors. To remove the topography associated with the gates, we polish the entire wafer using chemical mechanical planarization (CMP) process, leaving 100nm of oxide above the polysilicon gates. The resulting surface is atomically smooth, which should preserve the intrinsic structure of the nanotube that we deposit later in the process. Next, we deposit and etch 50nm of TiN to form the source/drain electrodes. To form ideal p-n diodes, we etch a deep trench between the source/drain contacts. The trench also helps to reduce the collection of secondary electrons. Finally, we lithographically define areas where we deposit the catalysts to grow the SWNTs. We grow the SWNTs at 900C using a catalytic chemical vapor deposition process in flowing CH$_4$ and H$_2$. We find both semiconducting and metallic devices. The resulting SWNTs are clean based on the ideal diode behavior we observe for semiconducting nanotubes. Fig. S1 shows the typical transfer curve from a metallic nanotube which is used in this study.

## 1.2. Current Analysis

We measure the current at both ends of the nanotube as the beam scans perpendicularly across it. The raw data are shown in Fig. S2A. We convert time to position using a scan rate of 270 nm/3.4 seconds. A settling time of about 3 seconds was used in order to allow the current to stabilize in the nanotube; there is a dark current present due to the inherent DC voltage offset between the current amplifiers that varies with time. We call this current $I_{DC}$. The DC current can be cancelled when we sum the source and drain currents; the result ($I_{SUM}$) shows an extremely low noise current with resolution in the fA range.

$I_{SUM}$ manifests an initial drop then rise in the current as the beam approached the nanotube. When the beam strikes the nanotube, the current drops sharply and recovers to baseline as shown in Fig. S2B. The beam is scanned in a line scan path such that the beam rescans the same line until the beam is blanked by the user. This results in many peaks for a given measurement. $I_{SUM}$ consists of two components: the dominant plasmon response ($I_P$) induced directly by the electron beam and a smaller secondary electron response ($I_{SE}$) that is due to the substrate at the bottom of the trench.

Secondary electrons are low energy electrons (< 50 eV) that are emitted from a sample in response to an incident high energy beam of electrons. These electrons are spatially emitted following cosine distribution relative the substrate normal (*25*). We show in Fig. S2B the typical $I_{SUM}$ profile. We clearly observe the $I_{SE}$ cosine profile as we show in Fig. S2C. Subtracting $I_{SE}$ from $I_{SUM}$ gives the desired signal profile $I_P$ in Fig. S2D. The subtracted signal should largely be featureless, except to vary slowly due to the offset voltage in the amplifiers. This confirms our assumption that $I_{SE}$ and $I_P$ contribute equally at each electrode.

In some cases, a slight offset is present between the peak floor to the left and right of the peak. For computational purposed, a line was drawn between the left and the right peak floors and then subtracted from the data to ensure the baseline on either side of the peak was zeroed out.

### *1.3. Setting an Astigmatism in Zeiss Ultra55 SEM*

In order to demonstrate the reconstruction technique, we created a beam with a known astigmatism to impart spatial features to the beam profile. Simply defocusing the beam will increase the measured FWHM or introduce a mesa-like profile (*4*). Although this proves the concept for one dimension, it says nothing about the other dimensions of the electron beam. Astigmatism occurs when there is uneven focusing power in the x and y coordinates of a focusing lens or when the beam itself enters the lens with varying cross-sectional intensity. Astigmatism in an electron beam has been discussed at length elsewhere (*28*).

In order to generate a consistent astigmatism for each measurement angle, we developed the procedure discussed below. We used a Zeiss Ultra 55 SEM, for which there are three controls that affect the beam shape: they are the "Focus" (F), "Stigmator X" ($X_{Stig}$) and "Stigmator Y" ($Y_{Stig}$). In practice, users frequently adjust these parameters to get the smallest beam spot possible and therefore the greatest resolution. In particular, $X_{Stig}$ and $Y_{Stig}$ are designed to correct the beam shape along their respective axis. By adjusting the relative correcting power, one can use the stigmators to construct a well-rounded or oval-shaped beam.

The measurement setup starts with determining the "in-focus" parameter values such that the observed image shows a sharp, high-resolution image. At this point, the beam is assumed to be finely focused and generally circular. Then, the $X_{Stig}$ and F values are offset by constants, $\delta X_{Stig}$ and $\delta F$ respectively, which were experimentally observed to produce the astigmatism shown in Fig. S3 D-G. In this way a known astigmatism can be produced for a given projection angle.

However, the parameter values that are needed for the smallest beam spot change depending on the electromagnetic field in the SEM chamber, among other things. Specifically, the field profile is affected by the relative positions of the device and the nanoprobes. This presents a challenge since we need to rotate the die and reposition the nanoprobes accordingly (See Fig. S4). This physical rearranging of objects near the beam changes the parameters needed to achieve the smallest beam spot. Consequently, using the same parameter values for each projection angle will not produce the desired astigmatism because the "in-focus" parameter values of one measurement would result in an "out-of-focus" condition for a different angle. Therefore, it is necessary to repeat the process outlined in Fig. S5 for each measurement angle.

To verify that this procedure results in a consistent astigmatism, highly-uniform gold nanoparticles of 30 diameter were dispersed on a "dummy device" which had no nanotube and utilized as an implicit indication of the beam shape. The procedure of rotating the stage and moving the probes around the chamber was carried out exactly as would be done for a line-scan measurement except only images of the gold nanoparticles were recorded. Figure S3 shows an image at various measurement angles demonstrating the reproducibility of the desired astigmatism at each angle. Table S1 shows the parameter values for the particle test and Table S2 shows the parameter values for the projection experiment shown in the main text.

*1.4. The Electron Beam Radon Transform*

The Radon transform is the basis for understanding the CT reconstruction from projection data and is readily understood by the illustration in Fig. 2A in the main text. The spatial profile obtained from our CNT measurement is equivalent to a single projection of a CT scan from a parallel x-ray source. While in a CT scan, contrast is the result of an x-ray signal that is attenuated by the varying physical density in the object of interest, in our measurement, contrast is the result of the varying current density of the electron beam as it gradually intercepts the nanotube. In either case, the Radon transform ($R(\rho,\theta)$) along an arbitrary line in the xy-plane, described in its normal representation as $\rho = x \cos\theta + y \sin\theta$, is given by:

$$R(\rho,\theta) = \iint_{-\infty}^{\infty} I(x,y)\delta(x\cos\theta + y\sin\theta - \rho)\,dx\,dy \qquad \text{Eq.1}$$

where $I(x,y)$ is the intensity map of the object being transformed (e.g. the electron beam distribution) and $\delta$ is the impulse function whose value is zero except when the argument is zero. In this way, only the values along the line will be included in the integral. If we fix theta ($\theta_k$) and compute $R(\rho,\theta)$ for values of $\rho$ which span $I(x,y)$, we can compute a single projection of $I(x,y)$ along the line we have defined. In our experiment, the line $\rho = x\cos\theta + y\sin\theta$ represents the position of the nanotube with respect to the centroid of the beam. Specifically, $\rho$ is the perpendicular distance from the centroid axis of the electron beam to the centroid axis of the carbon nanotube. By varying value $\theta$ and spanning $I(x,y)$ with $\rho$, we obtain several projections of the electron beam. The result of the Radon transform is typically displayed in what is called a 'sinogram'. The sinogram is a pseudo-graphical image where the y-axis is the distance along a projection and the x-axis is the projection angle The sinogram shows a collection of overlapping sinusoidal representations of objects in the xy plane.

Basic reconstruction from projection data involves back-projection of each of the projections across the image plane. For a single back projection at a given angle ($\theta$), we can write the following:

$$I_\theta(x,y) = R(x\cos\theta + y\sin\theta, \theta) \qquad \text{Eq.2}$$

If we integrate over many back-projections, we obtain a reconstructed image:

$$I'(x,y) = \int_0^\infty I_\theta(x,y)\,d\theta \qquad \text{Eq.3}$$

An issue associated with this simple reconstruction scheme is that if the data is not properly filtered, the reconstruction will be plagued by blurry, cloudy artifacts near the center as shown by Fig. S7A. In order to account for this, a bandlimited ramp filter, known as the 'Ram-Lak' filter (*29*), is applied in the Fourier domain. The filter is shown in Fig. S7 B. This process involves computing the 1-D Fourier transform of each projection, multiplying each Fourier transform by the 'Ram-Lak' filter, obtaining the inverse 1-D Fourier transform of each of the filtered projections, and back projecting the resulting 1-D inverse transforms to obtain the filtered reconstruction. Mathematically:

$$I'(x,y) = \int_0^\pi \left[\int_{-\infty}^{\infty} |\omega|\, \Gamma(\omega,\theta) e^{j2\pi\omega\rho} d\omega\right]_{\rho=x\cos\theta+y\sin\theta} d\theta \qquad \text{Eq.4}$$

where $\omega$ is the frequency term, $|\omega|$ is the Ram-Lak filter and $\Gamma(\omega,\theta)$ is the 1-D Fourier transform of a projection with respect to $\rho$. Figure S7 C and A show the contrasting filtered and non-filtered results of our measurements respectively. This mathematical formulation is based on the description of the Radon transform and image reconstruction presented at length elsewhere (*30*). While the formulation presented here is sufficient to understand the process, in practice a discrete implementation of the functions presented is necessary along with interpolation of the data to fit positionally in the reconstructed grid space.

A complication to our process is the fact that little can be done to ensure that the carbon nanotube is located centrally in the line scan profile. An assumption that is built into the previous discussion is that the object in question does not move around in space as the projection angle changes but is fixed at the origin. In our experimental setup, this is virtually impossible to achieve as our SEM stage lacks the precision to rotate and keep the same field of view within 1 nm. Therefore, a data processing approach was taken where the projections were shifted with subpixel translation to the center of gravity (COG). We define the COG of a curve as the point at which the integral to the left of the COG is equal to the integral to the right of COG. Each projection was shifted to its respective COG to compensate for the inevitable sample drift when rotating the nanotube.

## 2. Supplementary Text

### *2.1. Carbon Nanotube Contamination*

As the carbon nanotube is scanned with the electron beam, contamination builds up on the surface of the nanotube. This results in broadening in the current dip in the line scan measurements therefore obscuring the true measurement of FWHM of the electron beam. The contamination rate is proportional to the electron dose that the CNT receives over time. Figure S6 shows an inferred rate of contamination based on the measured FWHM of the current dip in the line scan measurements and the quantity of time that the CNT has been exposed to the electron beam.

According to the Nyquist Theorem, in order to measure a given frequency (f), the sampling rate (s) must be at least half the frequency. In spatial terms, this is the equivalent of the detector being at least half the size of the detail of interest. Kataura plot analysis on these devices indicates that the starting diameter of the CNTs is anywhere from 1.5 ~ 1.9 nm (*23, 31*) which shows that we are well within the Nyquist detection limit.

Nevertheless, we took care to minimize the contamination of the nanotube in the course of our measurements. Contamination build-up is the primary constraint influencing the number of projections chosen for the reconstruction. Ideally, we would capture many more projections to improve the accuracy. However, carbon contaminants will cause the CNT to swell to the point where the data is not useful for reconstruction. In this way, our measurement is somewhat limited by the rate of contamination. If we are going to beat this limit, a more complex model may be developed which accounts for the size of the nanotube with respect to electron dose over time. Alternatively, an ultra-high vacuum SEM chamber equipped with a heated stage may be used order to limit the contamination rate.

It was determined that a minimum of 10 measurements at 18-degree increments should be used in order to obtain a qualitatively good reconstruction. We performed all the measurements for the Radon transform in under 207 seconds of beam exposer therefore limiting the carbon contamination to a more useful range as shown in Fig. S6.

### *2.2. Gate Biasing Effects*

The buried split gates allow for biasing semiconducting nanotubes into a range of doped junction configurations (*24*). For completeness, we altered the gate biasing in various configurations of metallic nanotubes to check if biasing the gates had any effect on the behavior of our measurement. Figure S8 shows two configurations. It can be observed that the device behaves in the same way for both gate configurations. That is, the measurement manifests an offset due to the collection of secondary electrons and the peak is in the same direction for both the source and the drain electrodes.

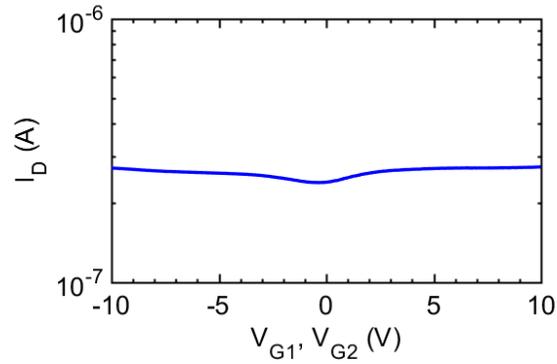

**Fig. S1.** The transfer curve of for a typical metallic nanotube device used in our study. The gates (G1 and G2 shown in Fig. 1.) are swept simultaneously from -10 V to +10 V while the drain current is monitored. The drain current does not vary significantly as the gates bias is swept which is characteristic of metallic devices. If this were a semiconducting nanotube, the current would vary by orders of magnitude.

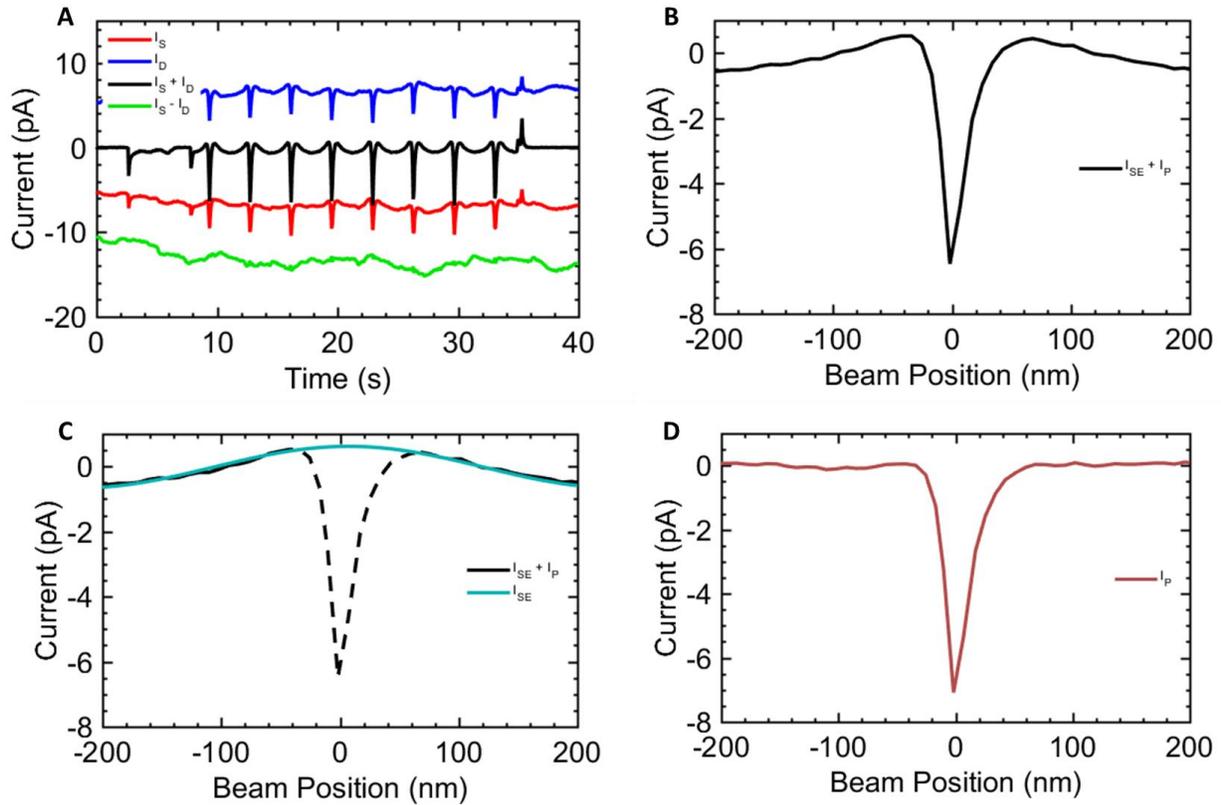

**Fig. S2.** Extracting the beam plasmon signal ($I_P$) from the raw current data. (**A**) The raw data collected for one angle in a series of projections. The electron beam scans a single line from left to right over the nanotube repeatedly generating multiple peaks until the user blanks the electron beam. (**B**) If we extract a single peak, we notice that there is a rise and fall to the background signal as the beam position changes leading up to and away from the nanotube respectively. We have determined that the carbon nanotube collects low energy secondary electrons (SEs) from the trench floor as they are generated in a cosine distribution relative to the substrate normal. We call this current $I_{SE}$. (**C**) Fitting a cosine curve to the background signal while ignoring the plasmon peak allows us to subtract the secondary signal from the total signal to isolate the plasmon signal in (**D**).

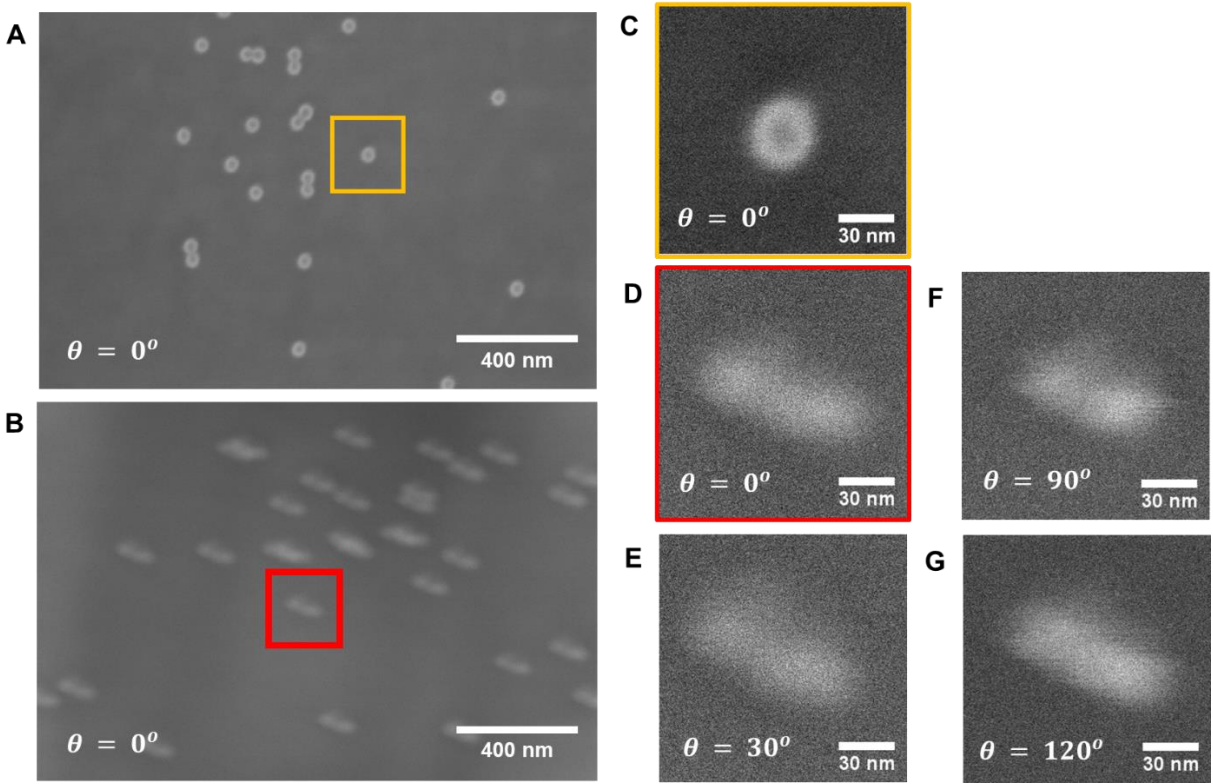

**Fig. S3.** Highly uniform, 30 nm gold nanoparticles were used as an indication of the astigmatism in the beam. Two different fields of particles are located on the sample and are imaged with (**A**) 'in-focus' and (**B**) 'astigmatism' parameters. A single particle is chosen from each field and shown in (**C**) and (**D**). The bright ring around the edge of the particle in (**C**) is known as edge brightening and is a well-known phenomenon in SEM. The astigmatism shown in (**D**-**G**) is applied to the beam for each angle (θ) according to the process laid out in Fig.S5. It can be observed that the desired astigmatism was created consistently across each angle although some vibration can be seen in the images particularly (**F**) and (**G**). Vibration is a limitation of our current set up and will need to be addressed in some fashion in the future.

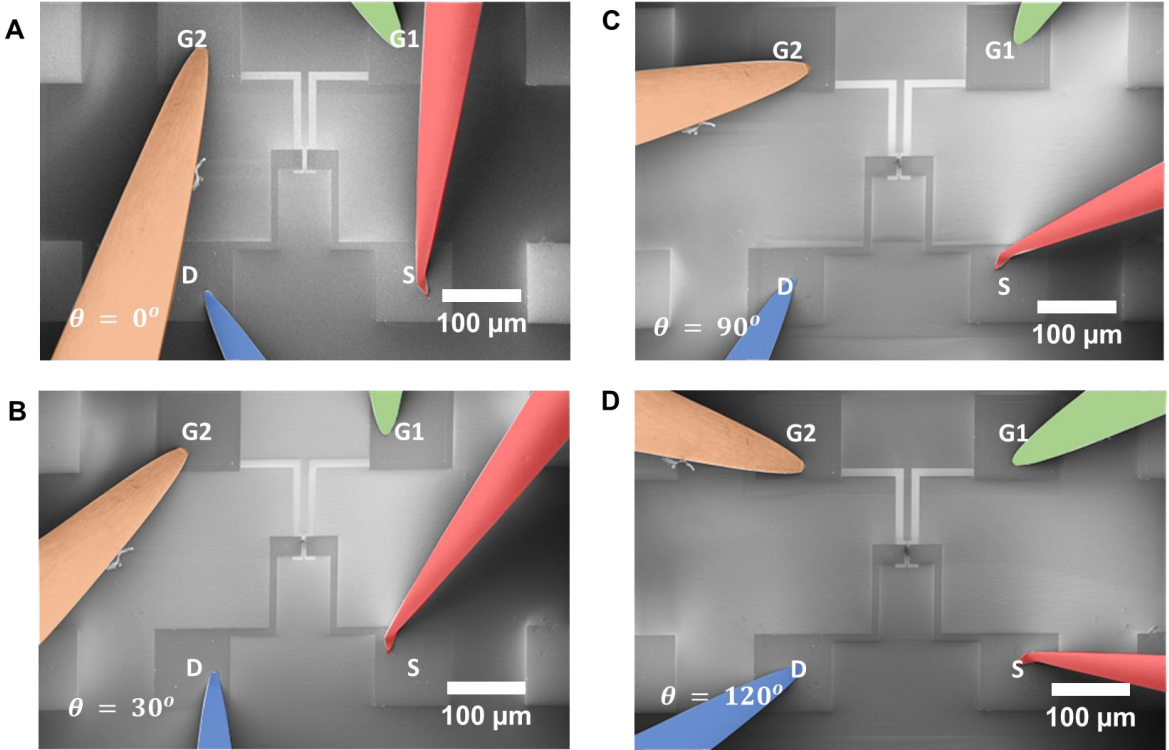

**Fig. S4.** A dummy device without a nanotube was coated with highly uniform, 30 nm gold nanoparticles and used to develop the procedure for rotating the device through the measurement angles. (**A**-**D**) The images show a low-mag view of the probes landed on the source (S), drain (D), gate 1 (G1) and gate 2 (G2) bond pads. The probes are color-coded to aid in showing how they are rotated with the sample for the different angle measurements. As the probes are rotated with the sample, the electromagnetic field in the sample chamber changes, resulting in different 'in-focus' parameters for each measurement angle.

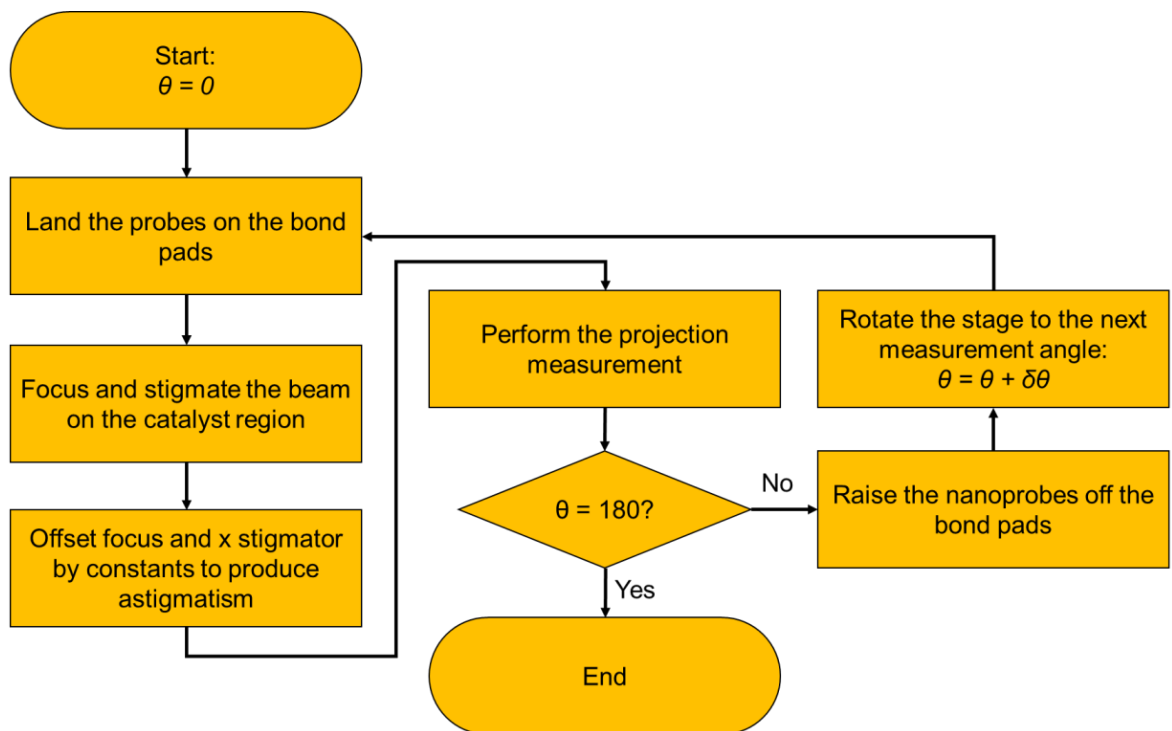

**Fig S5.** Process flow for collecting the projection data while maintaining a consistent astigmatism in the beam throughout the measurements.

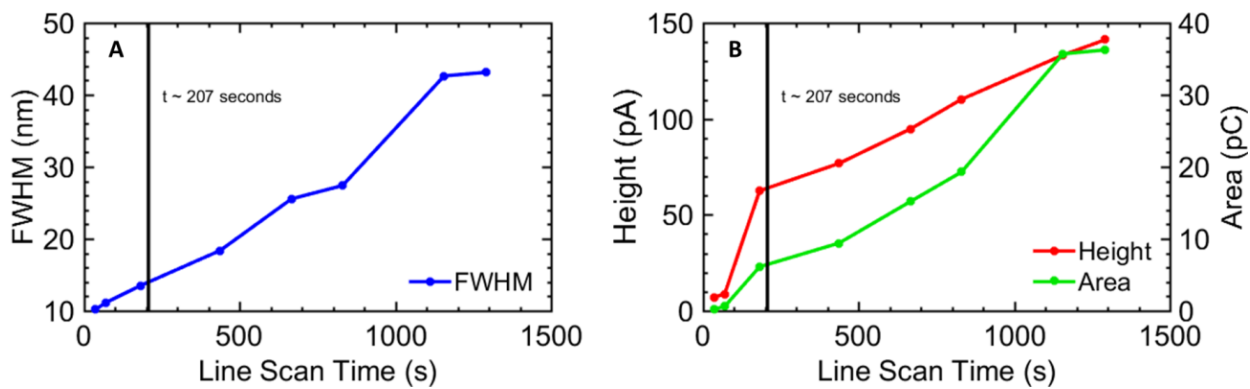

**Fig. S6.** The swelling effect of contamination build-up on the nanotube. (**A**) The apparent FWHM of the current peaks changes over time due to the contamination. (**B**) Height and the area of the peaks increased over time as well indicating that the contamination built up on all sides of the nanotube. The 'line scan time' is the total accumulated time that the device was exposed to the electron beam in line scanning mode. These curves were established on a sacrificial device and used as a parameter guide for how many measurements we would be able to take before the nanotube swelled significantly. All our measurements for the data presented in this paper were taken prior to t ~ 207 seconds.

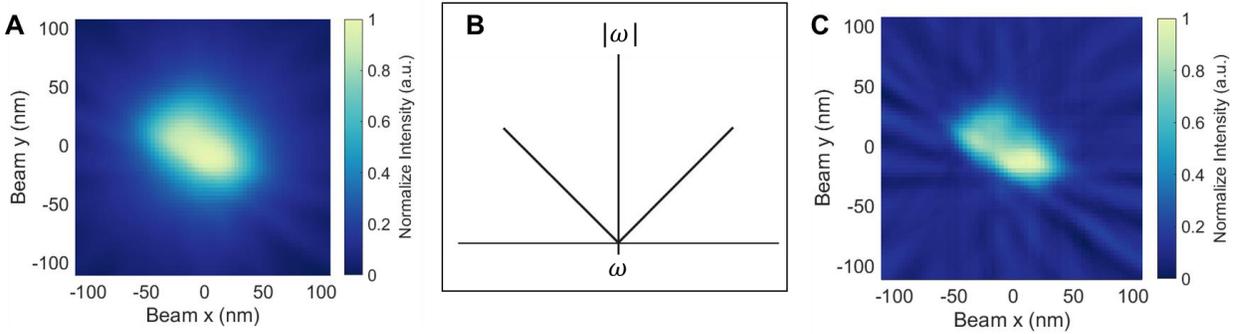

**Fig. S7.** The reconstruction of the beam profile from the projection data shown in Fig. 4. (**A**) Shows reconstruction of the beam by simple back-projection. It is known that back-projection in itself is not sufficient to adequately reconstruct the full beam profile. The projection data becomes oversampled in the center of the reconstructing leading to blurring. (**B**) The band limited ramp filter used to filter the projection data prior to back-projection. This is implemented according to Eq. 4 resulting in (**C**), the filtered reconstruction of the beam. This view is the top-down view of the same beam profile shown in Fig. 4.

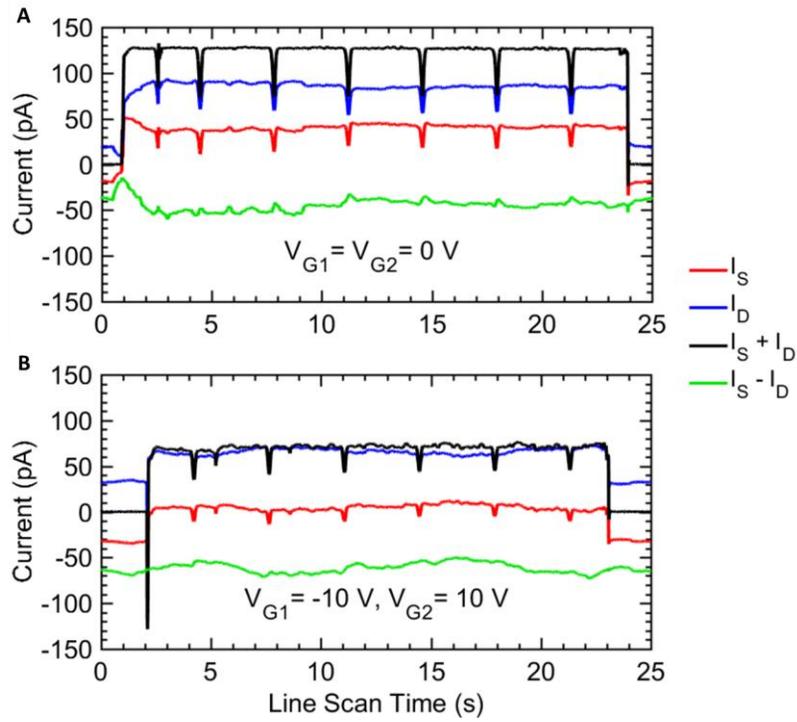

**Fig. S8.** Measurements on a device with (**A**) zero gate bias and (**B**) PN configuration to show that it behaves similarly in both conditions.

| Relative Stage Angle (Deg.) | In-Focus Parameters | | Desired Astigmatism Parameters | |
|---|---|---|---|---|
| | F (mm) | $X_{Stig}$ (%) | $F + \delta F$ (mm) | $X_{Stig} + \delta X_{Stig}$ (%) |
| 0.0 | 7.113 | -13 | 7.116 | -18 |
| 30.0 | 7.110 | -18 | 7.113 | -23 |
| 90.0 | 7.106 | -26 | 7.109 | -31 |
| 120.0 | 7.113 | -29 | 7.116 | -34 |

**Table S1:** SEM imaging parameters for the dummy die with highly uniform 30nm gold particles. The desired beam spot was accomplished by offsetting the 'In-Focus' parameters by constants for each measurement.

| RelativeStage Angle (Deg.) | In-Focus Parameters | | Desired Astigmatism Parameters | |
|---|---|---|---|---|
| | F (mm) | $X_{Stig}$ (%) | F + δF (mm) | $X_{Stig} + δX_{Stig}$ (%) |
| 0 | 7.024 | -36 | 7.027 | -41 |
| 18 | 7.024 | -35 | 7.027 | -40 |
| 36 | 7.026 | -34 | 7.029 | -39 |
| 54 | 7.028 | -31 | 7.031 | -36 |
| 72 | 7.029 | -25 | 7.032 | -30 |
| 90 | 7.030 | -23 | 7.033 | -28 |
| 108 | 7.031 | -20 | 7.034 | -25 |
| 126 | 7.033 | -17 | 7.036 | -22 |
| 144 | 7.035 | -14 | 7.038 | -19 |
| 162 | 7.036 | -11 | 7.039 | -16 |
| 180 | 7.040 | -10 | 7.043 | -15 |

**Table S2:** SEM parameters for the line scan, projection data shown in the main text (Fig.3.). The desired beam spot was accomplished by offsetting the 'In-Focus' parameters by constants for each measurement.